# USING THE CASE STUDY METHOD IN TEACHING COLLEGE PHYSICS

Lior M. Burko[1],
School of Science and Technology,
Georgia Gwinnett College, Lawrenceville, GA 30044
And
Department of Physics, Chemistry, and Mathematics,
Alabama A&M University, Normal, AL 35372

The case-study teaching method has a long history (starting at least with Socrates), and wide current use in business schools, medical schools, law schools, and a variety of other disciplines. However, relatively little use is made of it in the physical sciences, specifically in physics or astronomy. The case-study method should be considered by physics faculty as part of the effort to transition the teaching of college physics from the traditional frontal-lecture format to other formats that enhance active student participation. In this paper we endeavor to interest physics instructors in the case-study method, and hope that it would also serve as a call for more instructors to produce cases that they use in their own classes and that can also be adopted by other instructors.

What is the case-study method? Essentially, cases are "stories with a message" whose intent is to educate (Herreid, 1997). There is no one single case-study method. The richness and versatility of this method extends far beyond the original Socratic interrogation, which was brilliantly featured in the movie and follow-up television series *The Paper Chase* by Professor Kingsfield. The different types of cases can be categorized according to who does the case analysis: individual assignments, lecture, discussion, or small group activities. The case style can then be chosen to fit with one or more of these types, e.g., public hearing, trial, or the Socratic method cases would be most appropriately conducted in a discussion format; journal article cases, problem-based learning (PBL), debate role playing, or team learning cases are best appropriate in small groups; and storytelling cases are best appropriate in the lecture format (Herreid, 1998). Other styles of cases relevant for the physics courses can include experimental case studies (Arion, Crosby, and Murphy, 2000), or numerical modeling or simulation cases.

One common misconception about the use of case studies in the teaching of the natural sciences is that unlike business or law cases "there is one right answer" in science. Indeed, a fact-driven subject matter that includes only closed-ended questions would make a poor candidate for an exciting lesson plan using the case studies method (e.g., "What is the freezing temperature of water at 1 ATM?" Say you present this question to your students, how can you follow *that* up with a stimulating discussion?!?). But then, it would make a poor candidate for an exciting lesson plan using *any* teaching method. One can, however, teach the facts of science also using open-ended problems, which often involve student engagement. For example, the questions of "Should we terraform Mars?," "Was the Federal government right in stopping public funding of SETI?," or "Was the BICEP2 team right in coming out with a press release on allegedly finding evidence for cosmic inflation?" do not have one correct answer. Students may benefit from discussion or debate cases that include diverging lines of reasoning that support opposing opinions. We use the first two examples as *public hearing* cases in our astrobiology course. Such open-ended science or science policy questions can serve as the starting point for cases that will teach the facts of the relevant science in addition to doing so in a potentially exciting, entertaining, and stimulating way for many students.

---
[1] burko@ggc.edu

We include the case studies method in parts of the introductory calculus-based and algebra-based physics courses at Georgia Gwinnett College, in addition to an entire introductory course on astrobiology – the science of life in the universe, which we taught also at the University of Alabama in Huntsville. An introductory astrobiology course, which is an inherently multi-disciplinary subject, has become in recent years a popular choice for the fulfillment of the general education science requirement in many colleges and universities, and is taken also by many students who do not major in a science discipline.

In what follows we give examples for some of the different types of cases that we have used. The various styles can accommodate learning units in many different courses, appropriate for virtually all lower-division physics or astronomy courses, and to some degree also higher-division or even graduate level courses. For a unit on energy and power that we taught in the calculus-based introductory Physics sequence (but which can just as well be used for the algebra-based course) we used a case titled *The "Garabed": A Case Study on Power and Energy*, which invokes the historical anecdote of the so-called "Garabed," an (inevitably failed) 1918 proposal to create a perpetual motion machine of the first kind. Our lesson plan uses the *Interrupted Case* method (Herreid, 1998), and includes reading the appropriate chapter in Robert Park's *Voodoo Science* (Park, 2000) broken into section, and is supplemented by contemporary newspaper stories (which we chose from local area newspapers). Each section is followed by questions that students discussed in small groups (of 3 or 4 members). The questions asked range from fact-based questions (Knowledge or Comprehension (Remembering or Understanding) questions according to Bloom's taxonomy (Bloom et al, 1956)) to high cognitive-level questions (Synthesis or Evaluations (Evaluating and Creating) questions in Bloom's taxonomy, and of course also intermediate level questions (Application and Analysis (Applying and Analyzing)). Some of the questions include just-in-time teaching, specifically using the internet, such as "The text refers to Pascal's wager. What is Pascal's wager?" (Knowledge question in Bloom's taxonomy), or "Explain and critique the suggestion by the author that Congress's passing of a special act was a manifestation of Pascal's wager" (Application and Evaluation question)."

Another format of cases that we use several times in our Astrobiology course involves fiction. Plato used this case format in his telling of Atlantis in *Timaeus* and in *Critias*. One variant is a *short story* case that, to the best of our knowledge, includes correct science only, with which the students are guided to learn the science behind the story. The other variant is a short story that intentionally includes both correct and incorrect science, and the students are tasked to identify which is which. An example for the former is a case titled *Sheer Dumb Luck: The Earth's Magnetosphere*, that tells the story of a hypothetical future crewed mission to the Moon, the first after the Apollo missions were stopped in 1972. We supplement the short story with a reading of Lewis (2004) and of Holman (2006). The main problem the astronauts encounter is a giant solar flare that would *inter alia* inevitably give the crew lethal dose of radiation. The astronauts do not have enough time to either abort the mission (return to Earth or at least below the Van Allen Belt), or land on the Moon and seek shelter. There is, however, something they can do, but they do not think of it until it is too late to do anything about it, and being saved becomes a matter of chance. This case includes much information about the history and future of lunar and space exploration, solar wind and space weather, and elements of planetary science, and is followed by questions that again span the Bloom taxonomy. Students work in small groups, answer questions that accompany the reading assignments, and then the entire forum discusses the answers provided by the various groups.

Another short story is included in a case titled *The Never Ending Contamination – Radioactivity* that uses a different context of radioactivity than that is most commonly relevant for astrobiology. Specifically, it tells the story of a possible radioactive contamination on Earth because of the anticipated detonation by terrorists of a dirty bomb in a densely populated urban area. Since the same principles of radioactivity are applied both in terrestrial radioactivity and in astrobiology applications, the different context is used in order to expand the applicability of the course's subject matter and avoid compartmentalization of the material, and allows the case to be used also in other physics, chemistry, or integrated science courses. The short story is supplemented by reading Susskind (2008) to discuss probability in greater detail, and includes questions for discussion in groups and later in the whole forum.

Other formats of fiction-based cases we have used include videos or movies. One such case we use includes the Carl Sagan novel *Contact* and the major movie featuring Jodie Foster by the same title. In this case students are tasked with comparing and contrasting the science as presented in the novel with the science presented in the movie.

A third format of fiction-based cases we use is that of a dialogue, that follows in essence Galileo's style in *Dialogue Concerning the Two Chief World Systems*. (whose role is loosely based on Salviati), Dave (Sagredo), and John (Simplicio) are all graduate students who debate a number of topics in dedicated cases: *The Prebiotic*

*Soup*, *What is Life?*, and *The Creation of Synthetic Life*. We present the dialogues in the form of an *interrupted staged reading* (students reading in front of the entire forum the lines of the characters), each section followed by questions and discussion in groups and the entire forum. Other case formats that we used in the astrobiology course included reading newspaper excerpts, debate, and a public hearing.

One criticism of the case study approach that is sometimes made is that it does not allow for quantitative learning, and therefore is not appropriate for quantitative-based courses for science majors. We strongly disagree with that opinion: First, in the biological sciences cases that include quantitative reasoning are very common (Herreid, Schiller, and Herreid, 2014). Second, in our own cases, many of the questions we ask involve quantitative reasoning. In the lesson based on the aforementioned case on radioactivity, e.g., we include quantitative questions such as "You are analyzing Moon rocks that contain small amount of uranium-238, which decays into lead-206 with a half life of about 4.5 billion years. (a) In one rock from the lunar highlands, you determine that 55% of the original uranium-238 remains; the other 45% decayed into lead-206. How old is the rock? (b) In a rock from the lunar maria, you find that 63% of the original uranium- 238 remains; the other 37% decayed into lead-206. Is this rock older or younger than the highlands rock? By how much?" Questions like this one allow the student to transfer learning from the context of a terrorist dirty-bomb threat to the context of planetary science relevant for astrobiology or for astronomy, or indeed for any other application of radioactive dating, and also learn to use the relevant quantitative reasoning to answer conventional questions. The case study method does not limit quantitative reasoning!

A significant body of literature on the assessment of case teaching and active learning strategies is beginning to accumulate (Johnson and Johnson, 1993; Lundberg et al 1999). Most importantly, as many case methods make use of small group instruction, much of the latter's benefits (Hake, 1998; Udovic et al, 2002) are naturally inherent to case-study methodology. Importantly, "the positive effects of small-group learning were significantly greater for members of under-represented populations (African Americans and Latinas/os)" (Springer et al, 1999). Combining the conclusions of (Lundberg et al 2002) and of (Springer et al, 1999), a case-study approach is specifically beneficial to under-represented ethnic and gender groups, and focusing a course on case studies may therefore contribute to enrollment and retention for members of these groups.

The use of case studies combined with team learning, as done in our astrobiology course, has been used in a technical scientific discipline (organic chemistry) by Dinan and Frydrychowski (1995), who reported greater material coverage and higher grades in a national exam than when the traditional lecture method was used. (See also Cliff and Wright (1996).)

Our own data are consistent with these findings. Attitudinal pre- and post-studies suggest greater student enthusiasm towards the case-study approach than towards the traditional lecture.

A major advantage of the case–study method is its innate flexibility and versatility, as is manifest in the many different case types (e.g., whole class discussion, small group methods, problem-based learning, the interrupted case method, the intimate debate method, conflict cases, team-based learning, etc) (Herreid 2007). Focusing a course on the case study approach is therefore not a limiting restriction, as the flexibility for developing and choosing among the difference case-study types allows selection of the most applicable type to the subject matter of each case and to the course's students. A collection of cases available for adoption by instructors is published by the National Center for Case Study Teaching in Science at [http://sciencecases.lib.buffalo.edu/cs/](http://sciencecases.lib.buffalo.edu/cs/) and any instructor may submit contributions for review and inclusion in the collection. Case studies from the history of physics are published by the Nuffield Foundation at [http://www.nuffieldfoundation.org/practical-physics/case-studies-history-physics](http://www.nuffieldfoundation.org/practical-physics/case-studies-history-physics).

This work was supported by NSF Grants DUE-0941327 and DUE-1300717. The author thanks Sandra Enger.